# Rapid Vapor-Assisted Solution Process of Metal-Organic Chalcogenides for High-Performance Light-Emitting Diodes

Sang-Hyun Chin[1,+], Daseul Lee[2,+], Donggyu Lee[1,+], Kwanghyun Chung[1,+], Eunjong Yoo[1], Tong-Il Kim[1], Su Hwan Lee[2], Sang Woo Bae[2], Young-Hoon Kim[2,*], and Yeonjin Yi[1,*]

[1]Department of Physics, Yonsei University, 50 Yonsei-ro, Seodaemun-gu, Seoul 03722, Republic of Korea.

[2]Department of Energy Engineering, Hanyang University, 222 Wangsimni-ro, Seongdong-gu, Seoul 04763, Republic of Korea.

**\*Corresponding authors**.

E-mail address: younghoonkim@hanyang.ac.kr, (Y.K.), yeonjin@yonsei.ac.kr (Y.Y.)

[+] These authors contributed equally to this work.

**In Brief**

We report a rapid vapor-assisted solution process to fabricate luminescent metal-organic chalcogenide (MOC) thin films under mild conditions. This method enables the first demonstration of MOC-based light-emitting diodes (MOCLEDs), achieving red electroluminescence at 633 nm with an external quantum efficiency approaching 0.1%. The scalable and substrate-friendly nature of this process overcomes key limitations of conventional MOC synthesis, offering a new platform for stable, tunable, and compositionally diverse emissive materials in optoelectronics.

**Highlights**

- Developed a rapid, low-temperature vapor-assisted solution process for MOC film
- Enabled the first demonstration of MOC-based light-emitting diodes (MOCLEDs)
- Achieved electroluminescence at 633 nm with EQE approaching 0.1%
- Overcame conventional MOC processing limitations via a vacuum-free, scalable method
- Demonstrated potential of MOCs as next-generation emitters for displays and lighting

**The Bigger Picture**

Metal-organic chalcogenides (MOCs) represent a promising class of semiconducting materials with inherent chemical robustness, cost-effectiveness, and design tunability. However, traditional MOC fabrication methods—based on vacuum-deposited metal films and prolonged thermal treatment—suffer from substrate damage and limited scalability, hindering integration into practical devices. In this study, we overcome these barriers by introducing a rapid, vapor-assisted solution process that produces luminescent MOC thin films under mild conditions. By applying this approach, we demonstrate the first MOC-based light-emitting diodes (MOCLEDs), paving the way for the development of stable, scalable, and compositionally diverse emissive materials for future display and lighting technologies. Our findings position MOCs as a compelling alternative to conventional emitters in optoelectronics, and our process offers a blueprint for expanding their real-world applications.

**Summary**

Metal-organic chalcogenides (MOCs), robust crystalline assemblies composed of coinage metals, chalcogens and organic ligands, are typically synthesized via prolonged, high-temperature tarnishing of vacuum-deposited metal films with organochalcogen precursors. The prolonged exposure to high temperatures and the necessity for direct vacuum deposition of silver can induce damage to the underlying films, posing significant challenges to the fabrication of optoelectronic devices, despite their cost-effectiveness and chemical robustness. This study introduces vapor-assisted solution processing, a novel chemical vapor deposition method, enabling remarkably rapid fabrication of luminescent MOC films. Furthermore, the first MOC-based light-emitting diodes (MOCLEDs) are realized, achieving an external quantum efficiency (EQE) approaching 0.1% and electroluminescence peaking at 633 nm. These results highlight the potential of MOCs as next-generation emitters for displays and solid-state lighting. This work offers a promising fabrication strategy and insights for advancing MOCLEDs and expanding their optoelectronic potential.

**Introduction**

In recent decades, significant research has focused on synthesizing single-layer or few-layered low-dimensional materials, such as graphene and transition metal dichalcogenides (TMDCs), for electronic and optoelectronic applications.[1] The introduction of heterogeneous elements or organic molecules on their surface tunes material properties, but harsh treatments often compromise the integrity of nanosheets.[2] Furthermore, such post-modification is typically inefficient, resulting in non-uniform atomic/organic group distribution and limiting practical applications. These materials also often lack well-defined structures, making understanding the relationship between structures and properties difficult. Low-dimensional halide perovskites have also been promising candidates for optoelectronics. Although three-dimensional perovskites with the general formula $ABX_3$ — where A is a monovalent cation, either an organic or an inorganic, B is a divalent heavy metal ion, and X is a halide anion — exhibit small exciton binding energies ($E_b$ ~50 meV) and considerable trap densities, introducing bulky organic cation (L) leads to the formation of quasi-two-dimensional perovskite phases ($L_2[ABX_3]_{(n-1)}BX_4$) with increased $E_b$ and limited traps, which is favorable for light-emitting devices.[3–10] However, halide perovskites suffer from limited intrinsic stability due to their ionic bond formation.[11]

A new class of low-dimensional materials, metal-organic chalcogenides (MOCs), first reported in 1991, addresses the above challenges.[12] Unlike well-known TMDCs and halide perovskites, MOCs exhibit a unique structure with a repeating low-dimensional unit of metal chalcogenide layer covalently bonded with organic functional groups, held together with van der Waals interaction. In 2018, Hohman *et al.* synthesized and examined the luminescent property and chemical robustness of two-dimensional silver benzene-selenolate ($[AgSePh]_\infty$), naming it "mithrene."[13] This synthesis involved pipetting diphenyl diselenide-in-toluene solution onto the aqueous silver(I) nitrate solution and harvesting the product over several days (3 to 10, depending on reactant concentration). The resulting AgSePh precipitates exhibited a sharp photoluminescence peak at 467 nm, but their morphology was unsuitable for thin-film optoelectronic devices. Recent studies by Tisdale *et al.* examined the luminescent properties of AgSePh from low temperature (5 K) to room temperature (RT).[14] AgSePh films with improved morphology were synthesized by tarnishing vacuum-deposited silver films (15 nm) with diphenyl diselenide vapor for 3 days. However, the photoluminescent quantum yield (PLQY) remained below 0.1 % at RT. The drawbacks of prolonged MOC synthesis and low PLQY can be overcome by optimizing the composition (coinage metal, chalcogen, organic ligand) and reaction pathways. Rapidly obtainable, highly luminescent MOC films with suitable morphology are promising for integrating light-emitting diodes (LEDs) and other optoelectronic applications.[12,15–19] A recent study reported a highly luminescent one-dimensional MOC, silver(I) 2-methyl ester benzenethiolate, with a PLQY of 22 %.[20] This synthesis involved reacting silver(I) oxide and methyl thiosalicylate ligand in methanol for 5 days. The resulting precipitates exhibit short-range Ag–O interactions, potentially enhancing luminescence by facilitating excitonic recombination or modifying electronic transitions to favor radiative decay. However, the electroluminescence of these materials has not been reported yet.

This report presents a vacuum-free synthetic method for MOC materials having one-dimensional wire structure, characterizes their luminescent properties, and demonstrates the first MOC-based LEDs (MOCLEDs) with an external quantum efficiency of 0.1%. MOC films are fabricated using a facile and rapid chemical vapor deposition (CVD) process, called vapor-assisted solution processing, previously employed in halide perovskite optoelectronics.[21,22] The two-step process involves spin-coating inorganic precursors followed by reacting to the resulting films with vaporized organic counterparts. A similar CVD method, utilizing vaporized organic ligands (benzeneselenol) and $O_2$ plasma-treated silver films, achieved MOC film fabrication within 1 hour and successful photodiode integration.[23,24] However, this method still relied on vacuum-deposited silver films, requiring longer deposition time and limiting direct coupling with the CVD process. To address this, solution-processable silver(I) nitrate and ambient-pressure sublimable methyl thiosalicylate ligand are employed in this study.

**Results and discussion**

**Figure 1A** illustrates the MOC fabrication reaction pathway. Spin-coated silver(I) nitrate films are enclosed in a glass petri dish with a drop of liquid-phase methyl thiosalicylate and heated (at 130 °C for 20 minutes) for the CVD process.

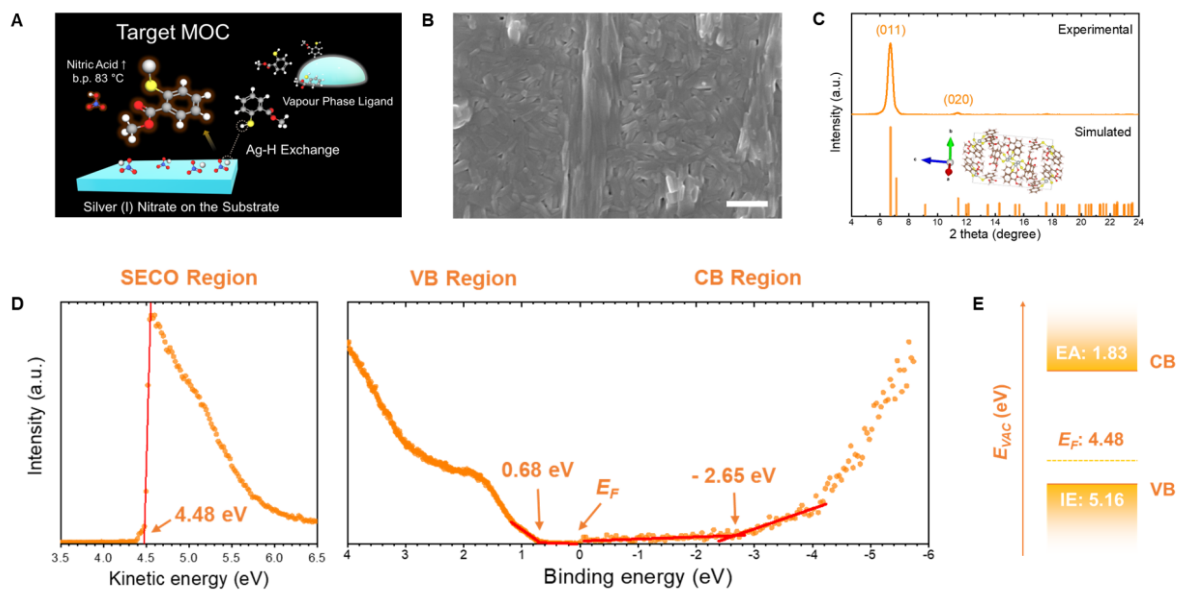

**Figure 1. Synthesis of Metal organic chalcogenide (MOC) films**. (A) Schematic illustration of chemical vapor deposition to fabricate MOC films. (B) Scanning electron microscopic image of MOC films (scale bar: 500 nm). (C) Experimental and simulated X-ray diffraction pattern of MOC films (inset: crystal structures of MOC, created with VESTA software based on crystallographic information files obtained CCDC-2212139). (D) Ultraviolet photoelectron spectroscopy (UPS) and Inverse photoemission spectroscopy (IPES) spectra of MOC presenting secondary cutoff (SECO), valence band (VB), and conduction band (CB) regions. (E) Energy-level diagrams deduced from the UPS-IPES spectra. ($E_{VAC}$ and $E_F$ denote vacuum level and Fermi level, respectively.)

The silver replaces the hydrogen in the thiol group, forming silver 2-methyl ester salicylate MOC solid and nitric acid. The nitric acid is released, due to its low boiling point (83 °C), upon opening the top of the CVD setup, leaving the MOC films as the target product as shown below.

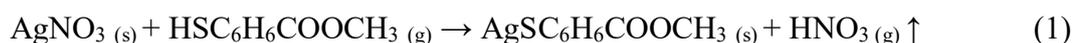

$$AgNO_{3\,(s)} + HSC_6H_6COOCH_{3\,(g)} \rightarrow AgSC_6H_6COOCH_{3\,(s)} + HNO_{3\,(g)} \uparrow \qquad (1)$$

**Figure 1B** shows the scanning electron microscopic (SEM) image of 100 nm-thick MOC films fabricated by spin-coating 0.1 M acetonitrile solution of silver(I) nitrate at 8000 rpm onto indium tin oxide (ITO) substrates, followed by CVD with 50 μl organic ligand for 20 minutes. The MOC films confirmed by SEM show pin-hole free, compact, and polycrystalline morphology, and long rod-like grain with the size of several hundreds of nanometers. X-ray diffractometry (XRD) in **Figure 1C** confirms the structural properties. The XRD pattern shows a sharp and intense peak near 6.7°, corresponding to the (011) plane, which is consistent with a previous report. However, an additional (020) peak is observed near 7.2°, indicating a different orientation of our CVD-grown MOC films.[20,25] Ultraviolet photoelectron spectroscopy (UPS) and inverse photoelectron spectroscopy (IPES) measurements show (**Figure 1D**) the electronic structure of the MOC films. The evaluated work function, ionization energy (IE), and electron affinity (EA) are summarized in **Figure 1E**. The combined UPS/IPES measurements indicate an electronic transport band gap ($E_g$) of 3.33 eV, close to the reported theoretical value of 3.5 eV, and strong p-type characteristics.[20]

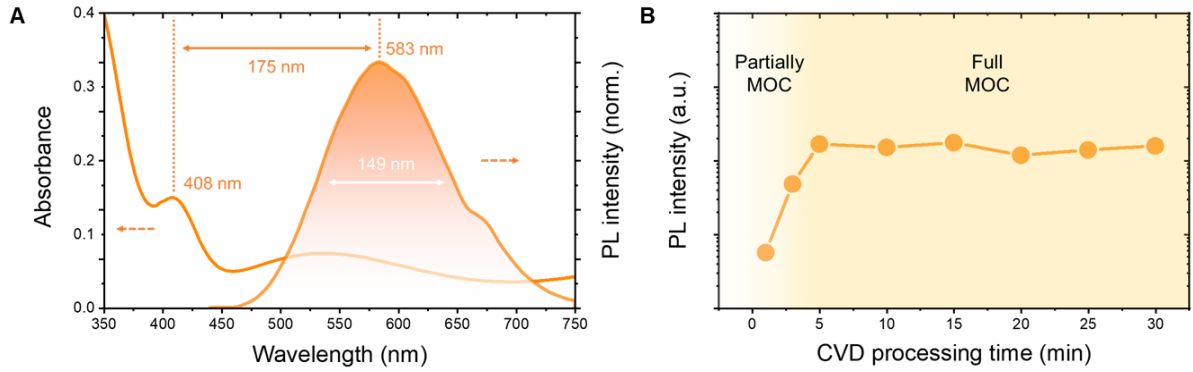

**Figure 2. Optical Property of CVD-grown MOC films.** (A) Absorbance and photoluminescence spectra of MOCs. (B) Peak PL intensity of MOC films upon process time.

Optical properties were investigated with absorption and photoluminescence (PL) spectra, as shown in **Figure 2A**. An excitonic transition peak ($E_{ext}$) centered at 408 nm (3.04 eV) is observed, followed by a monotonic increase in inter-band absorption. Considering the measured transport band gap of 3.33 eV, the exciton binding energy ($E_b$) can be obtained as 290 meV by the equation below.[26]

$$E_b = E_g - E_{exc} = (IE - EA) - E_{exc} \qquad (2)$$

The MOC films exhibit an external PLQY of 18.5% and an internal PLQY of 37.5% (Supporting Information, **Figure S1**). Furthermore, the PL spectrum displays a broad emission, with a full width at half maximum (FWHM) of 149 nm centered at 583 nm, corresponding to a Stokes shift of 175 nm. To assess the rapidity of CVD process, peak PL intensities of MOC films were measured as a function of CVD processing time (**Figure 2B**). The results indicate that silver(I) nitrate to MOC conversion occurs within 5 minutes, after which luminance stabilizes. It demonstrates that the CVD process introduced in this study is both remarkably effective and rapid. The promising optical characteristics and rapid conversion motivate the integration of MOC into LED devices, to evaluate their electroluminescent properties and overall performance.

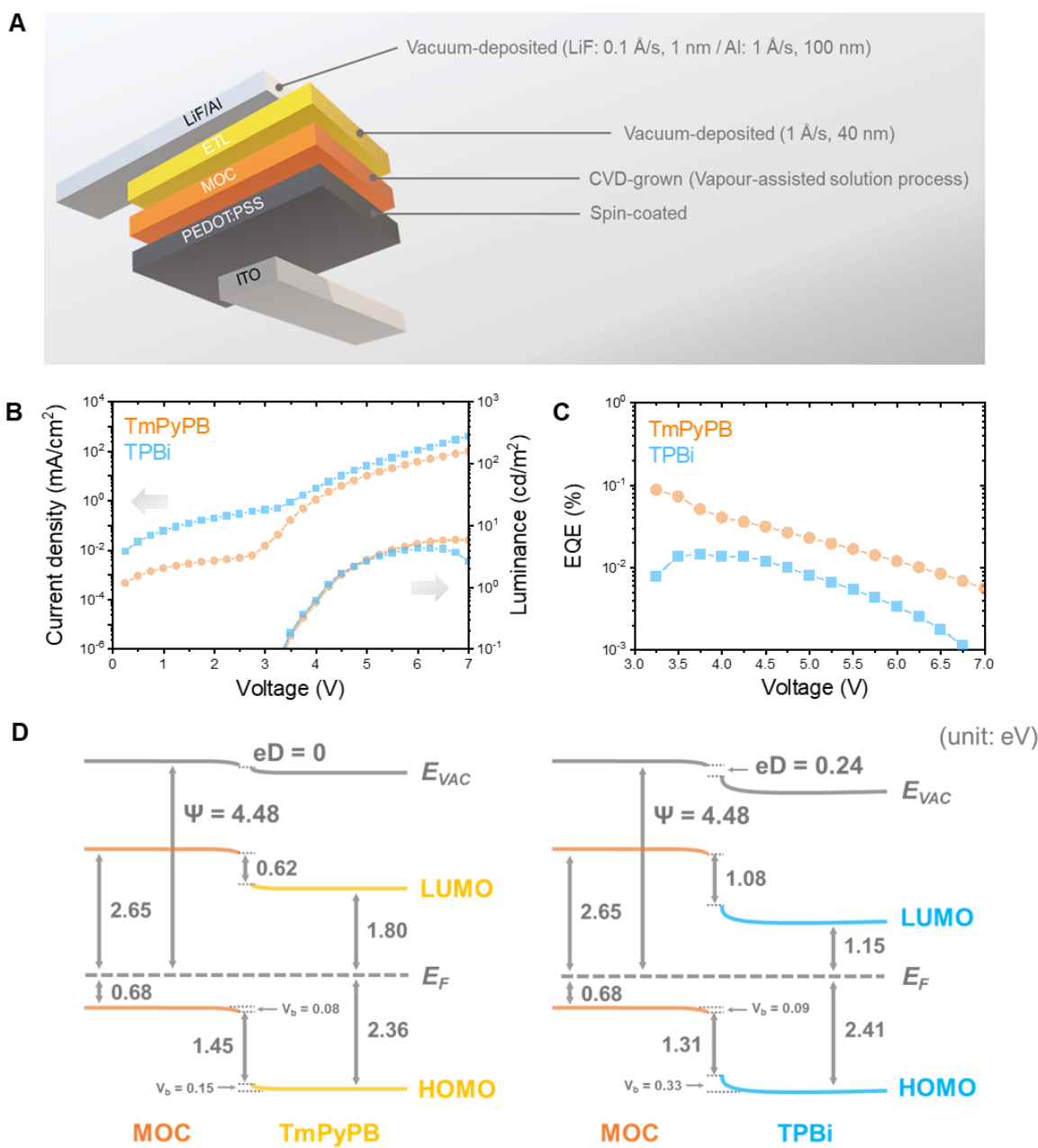

**Figure 3. MOC-based Light-Emitting Diodes (MOCLEDs).** (A) Device structure of MOCLEDs. (B) Current density-voltage-luminance characteristics of MOCLEDs comprising TmPyPB and TPBi electron transport layers. (C) External quantum efficiency (EQE) of MOCLEDs. Energy level diagram of (D) MOC-TmPyPB and MOC-TPBi interface. $E_{VAC}$, $E_F$, $V_b$, eD, and Ψ denote the vacuum level, Fermi level, band bending, interface dipole, and work function, respectively.

**Figure 3A** schematically illustrates MOC-based LEDs (MOCLEDs). The fabrication process begins with spin-coating of poly(3,4-ethylenedioxythiophene) polystyrene sulfonate (PEDOT:PSS) onto UV-ozone-treated ITO substrates. Subsequently, a 0.05 M silver(I) nitrate solution in acetonitrile is spin-coated at 8000 rpm. Films are then exposed to methyl thiosalicylate ligand via CVD in a nitrogen-filled glovebox. After the MOC growth, samples are transferred to a glovebox-integrated vacuum thermal evaporator for device completion. 1,3,5-tri(m-pyridin-3-ylphenyl) benzene (TmPyPB) or 2,2',2"-(1,3,5-benzinetriyl)-tris(1-phenyl-1-H-benzimidazole) (TPBi), having the dual role of electron transport and hole-blocking layers, are deposited to a 40 nm thickness for comparative analysis. Finally, LiF (1 nm) and an Al (100 nm) electrode are deposited, defining a 5 mm$^2$ active area. As a result, the electroluminescence (EL) spectra of the MOCLEDs, as shown in Supporting Information **Figure S2**, exhibit peak emissions at 618 nm for the TPBi-based device and 633 nm for the TmPyPB-based counterpart, both of which are redshifted relative to the photoluminescence peak at 583 nm.

**Figure 3B** depicts the current density-voltage-luminance (J-V-L) characteristics, revealing a significantly higher current density in the TPBi-based MOCLED compared to the TmPyPB-based MOCLED at equivalent voltages. Conversely, the TmPyPB-based MOCLED exhibits slightly higher maximum luminance than does TPBi-based MOCLED despite its lower current density, indicating a greater proportion of injected carriers undergoing radiative recombination. This correlation between current density and luminance is further evidenced in **Figure 3C**, where the TmPyPB-based MOCLED demonstrates a substantially higher EQE for all voltage ranges. The higher EQE (0.09 % for the TmPyPB-based MOCLED and 0.014 % for the TPBi-based counterpart) directly results from the lower current density and higher luminance in the TmPyPB-based device. To elucidate the performance variations attributed to electron transport materials (ETMs), *in situ* photoelectron spectroscopy is employed.[27–29]

Two distinct ETMs are thermally deposited in a vacuum and directly transferred to the spectroscopy chamber without breaking the vacuum (Supporting information **Figure S3**). Sequential depositions and measurements are repeated (ETMs with thicknesses of 2, 5, 10, 20,

50, 100, and 200 Å, Supporting information **Figure S4**) to analyze the MOC/ETM interface, including charge injection/blocking barriers, band bending ($V_b$) and interface dipole (eD). In addition, the electronic structure of ETMs is investigated prior to the interface experiment to understand the pristine properties of ETMs (Supporting information **Figure S5**). **Figure 3D** illustrates the energy level diagrams of the MOC/TmPyPB and MOC/TPBi interfaces. The total work function ($\Psi$) shift consists of the eD and $V_b$, with the relation of $\Delta\Psi = -(eD + V_b)$.[30] A significant eD (0.24 eV) is observed at the MOC/TPBi interface, whereas the MOC/TmPyPB interface shows no detectable eD. Both TmPyPB and TPBi provide considerable hole-blocking barriers against MOC (1.45 and 1.31 eV, respectively), with TmPyPB more strongly confining holes to enhance electron-hole recombination. Notably, the electron injection barrier from TmPyPB to MOC (0.62 eV) is lower than that from TPBi to MOC (1.08 eV), suggesting improved device efficiency with TmPyPB as an ETM.[31] While the results demonstrate promising performance, suboptimal charge carrier management remains a limiting factor in device efficiency. In addition, it is essential for the research community to explore effective doping strategies and defect passivation techniques.[32,33] Efficiency losses due to anti-site defects or vacancies in MOCs could significantly impact charge transport and recombination dynamics, necessitating further studies. Overcoming these challenges will be crucial for unlocking the full potential of MOCLEDs and facilitating their practical applications.

**Conclusions**

This study presents a facile chemical vapor deposition (CVD) method, for fabricating highly luminescent metal-organic chalcogenide (MOC) films and demonstrates their integration into the first MOC-based light-emitting diodes (MOCLEDs). The innovative CVD process introduced in this report enables MOC film fabrication within 5 minutes, significantly faster than previous methods requiring several hours to days. Electroluminescence and photoelectron spectroscopy studies reveal that selecting the electron transport layer (ETL) impacts device performance critically. MOCLEDs employing TmPyPB as an ETL exhibit a sixfold enhancement in external quantum efficiency (EQE), reaching 0.1 %, attributed to the improved hole blocking and electron injection. These findings highlight the potential of MOCs as next-generation luminescent materials for optoelectronic applications. The demonstrated CVD process enables scalable and rapid fabrication, paving the way for future studies to further optimize the composition, structural control, and charge transport properties of MOCs. By refining these aspects, MOCLEDs could emerge as competitive candidates for high-efficiency light-emitting devices.

**Methods**

*Materials*

Methyl thiosalicylate (97 %), acetonitrile (99.8%, anhydrous) and silver(I) nitrate (99.0 %) were purchased from Sigma Aldrich. 2,2′,2″-(1,3,5-benzinetriyl)-tris(1-phenyl-1-H-benzimidazole) (TPBi) and 1,3,5-Tris(3-pyridyl-3-phenyl)benzene (TmPyPB) were purchased from EM Index. LiF and aluminium were purchased from iTASCO.

*Synthesis of MOCs*

Silver(I) nitrate solutions (0.1 M for characterization, 0.05 M for devices) in acetonitrile were spin-coated onto substrates at 8000 rpm. The resulting films were placed in a glass petri dish with 50 µL of methyl thiosalicylate ligand and heated on a hot plate at 130 °C.

*Fabrication of MOCLEDs*

Patterned indium tin oxide (ITO) glasses were cleaned by sonication in acetone and 2-isopropanol for 15 minutes each and boiled in 2-isopropanol. After cleaning, the glasses were subjected to UV–ozone treatment for 10 minutes. Subsequently, a poly(3,4-ethylenedioxythiophene)-poly(styrenesulfonate) (PEDOT:PSS) hole-injection layer was spin-coated onto the ITO patterned glasses and baked at 150 °C for 1 hour. MOC films were fabricated onto the PEDOT:PSS described previously. The resulting films were transferred to a glovebox-integrated vacuum thermal evaporator, where the electron transport layer (TPBi or TmPyPB) (40 nm) and LiF (1 nm)/aluminum (100 nm) were sequentially deposited via thermal evaporation under high vacuum (<$10^{-7}$ Torr).

*Characterization*

The surface morphology of MOC thin films was characterized using field-emission scanning electron microscopy (FE-SEM, 7610F-Plus, JEOL). High-resolution XRD data, including $\theta-2\theta$ scans, were collected diffractometer in a parallel beam geometry using Cu Kα

($\lambda$ = 1.5406 Å, four-bounced Ge (220) monochromatized beam) X-ray tube operated at 45 kV/30 mA (Smartlab, Rigaku). MOC absorbance was measured using UV-Vis spectrophotometer (V-650, JASCO). Steady-state photoluminescence was measured using a spectrometer (FS-2, SCINCO), and photoluminescent quantum yields were evaluated using an integrating sphere-coupled spectrometer (FP-8550, JASCO). Ultraviolet photoelectron spectroscopy (UPS) was performed using a SPECS PHOIBOS 150 hemisphere analyser and a He I$_\alpha$ ($hv$ = 21.22 eV) UV discharge lamp. The secondary electron cutoff (SECO) was obtained with a − 5 V bias during UPS measurements. Inverse photoelectron spectroscopy (IPES) was conducted in isochromat mode with a low-energy electron gun and a bandpass filter (SrF$_2$-NaCl combination for 9.5 eV pass energy). The energy reference for both UPS and IPES was calibrated using clean Au. Current density-voltage-luminance (J-V-L) characteristics of MOCLEDs were measured using a Keithley 2400 source measurement unit and a Minolta CS2000 spectroradiometer.

**Acknowledgements**
This work was supported by the National Research Foundation of Korea (Grant No. RS-2024-00335481 and No. RS-2023-00276411) and YAS Company.

**Contributions**
Y.Y. and Y.H.K. directed the project and conceived the idea. S.H.C. designed the synthetic process. S.H.C., K.C., E.Y., and T.K. investigated structural and optical properties. S.H.C. and D.G.L. investigated the electronic structure of materials and interfaces. S.H.C. and D.S.L. fabricated devices. D.S.L., S.L., and S.B. evaluated the devices. S.H.C. drafted the paper, and all co-authors reviewed the paper.

**References**

1. Tan, C. *et al.* Recent advances in ultrathin two-dimensional nanomaterials. Chem. Rev. 117, 6225–6331 (2017).
2. Voiry, D. *et al.* Covalent functionalization of monolayered transition metal dichalcogenides by phase engineering. Nat. Chem. 7, 45–49 (2015).
3. Chin, S.-H. Perovskite multiple quantum wells: toward artificial construction and lasing. Discover Appl. Sci. 6, 396 (2024).
4. Fakharuddin, A. *et al.* Perovskite light-emitting diodes. Nat. Electron. 5, 203–216 (2022).
5. Chin, S.-H. Artificial perovskite multiple quantum well optoelectronics. J. Opt. Photonics Res. (2024). https://doi.org/10.47852/bonviewJOPR42022936
6. Kim, G. Y. *et al.* Controlled phase distribution of quasi-2D perovskite enables improved electroluminescence. J. Phys. Energy 6, 035002 (2024).
7. Kim, Y.-H., Cho, H. & Lee, T.-W. Metal halide perovskite light emitters. Proc. Natl. Acad. Sci. U.S.A. 113, 11694–11702 (2016).
8. Kim, Y.-H. *et al.* Exploiting the full advantages of colloidal perovskite nanocrystals for large-area efficient light-emitting diodes. Nat. Nanotechnol. 17, 590–597 (2022).
9. Kim, Y.-H. *et al.* Comprehensive defect suppression in perovskite nanocrystals for high-efficiency light-emitting diodes. Nat. Photonics 15, 148–155 (2021).
10. Cho, H. *et al.* Overcoming the electroluminescence efficiency limitations of perovskite light-emitting diodes. Science 350, 1222–1225 (2015).
11. Park, B.-W. & Seok, S. I. Intrinsic instability of inorganic–organic hybrid halide perovskite materials. Adv. Mater. 31, 1805337 (2019).
12. Dance, I. G., Fisher, K. J., Herath Banda, R. M. & Scudder, M. L. Layered structure of crystalline compounds AgSR. Inorg. Chem. 30, 183–187 (1991).
13. Schriber, E. A. *et al.* Mithrene is a self-assembling robustly blue luminescent metal–organic chalcogenolate assembly for 2D optoelectronic applications. ACS Appl. Nano Mater. 1, 3498–3508 (2018).
14. Lee, W. S. *et al.* Light emission in 2D silver phenylchalcogenolates. ACS Nano 16, 20318–20328 (2022).
15. Hernandez Oendra, A. C. *et al.* Tunable synthesis of metal–organic chalcogenide semiconductor nanocrystals. Chem. Mater. 35, 9390–9398 (2023).
16. Lee, W. S. *et al.* Mixed-chalcogen 2D silver phenylchalcogenides ($AgE_{1-x}E_xPh$; E = S, Se, Te). ACS Nano 18, 35066–35074 (2024).
17. Sakurada, T. *et al.* 1D hybrid semiconductor silver 2,6-difluorophenylselenolate. J. Am. Chem. Soc. 145, 5183–5190 (2023).
18. Hawila, S. *et al.* Tuning the 1D–2D dimensionality upon ligand exchange in silver thiolate coordination polymers with photoemission switching. J. Mater. Chem. B 11, 3979–3984 (2023).
19. Chin, S., Mardegan, L., Palazon, F., Sessolo, M. & Bolink, H. J. Dimensionality controls anion intermixing in electroluminescent perovskite heterojunctions. ACS Photonics 9, 2483–2488 (2022).

20. Aleksich, M. *et al.* Ligand-mediated quantum yield enhancement in 1D silver organothiolate metal–organic chalcogenolates. Adv. Funct. Mater. 35, 2414914 (2025).
21. Chen, Q. *et al.* Planar heterojunction perovskite solar cells via vapor-assisted solution process. J. Am. Chem. Soc. 136, 622–625 (2014).
22. Chin, S.-H. *et al.* Realizing a highly luminescent perovskite thin film by controlling the grain size and crystallinity through solvent vapour annealing. Nanoscale 11, 5861–5867 (2019).
23. Maserati, L., Pecorario, S., Prato, M. & Caironi, M. Understanding the synthetic pathway to large-area, high-quality [AgSePh]∞ nanocrystal films. J. Phys. Chem. C 124, 22845–22852 (2020).
24. Maserati, L. *et al.* Photo-electrical properties of 2D quantum confined metal-organic chalcogenide nanocrystal films. Nanoscale 13, 233–241 (2021).
25. Abdallah, A. *et al.* Luminescent and sustainable d10 coinage metal thiolate coordination polymers for high-temperature optical sensing. iScience 26, 106958 (2023).
26. Kim, B. *et al.* Tuning electronic structure and carrier transport properties through crystal orientation control in two-dimensional Dion-Jacobson phase perovskites. Nano Convergence 12, 1 (2025).
27. Jung, K. *et al.* Energy level alignments at the interface of N,N'-bis-(1-naphthyl)-N,N'-diphenyl-1,1'-biphenyl-4,4'-diamine (NPB)/Ag-doped $In_2O_3$ and NPB/Sn-doped $In_2O_3$. Appl. Surf. Sci. 387, 625–630 (2016).
28. Jung, K. *et al.* Elucidation of hole transport mechanism in efficient energy cascade organic photovoltaics using triple donor system. Appl. Surf. Sci. 576, 151866 (2022).
29. Schlaf, R. *et al.* Determination of interface dipole and band bending at the Ag/Tris(8-hydroxyquinolinato) gallium organic Schottky contact by ultraviolet photoemission spectroscopy. Surf. Sci. 450, 142–155 (2000).
30. Lüth, H. Solid Surfaces, Interfaces and Thin Films. Graduate Texts in Physics (Springer, 2010).
31. Sun, Y., Chen, S., Huang, J. Y., Wu, Y. R. & Greenham, N. C. Device physics of perovskite light-emitting diodes. Appl. Phys. Rev. 11, 011301 (2024).
32. Jiang, N. *et al.* Defects in lead halide perovskite light-emitting diodes under electric field: from behavior to passivation strategies. Nanoscale 16, 3838–3880 (2024).
33. Chin, S.-H. & Lee, J.-W. Towards the optimal interstitial doping for halide perovskites. Nano Res. Energy 2, 20230012 (2023).

# Graphical Abstract

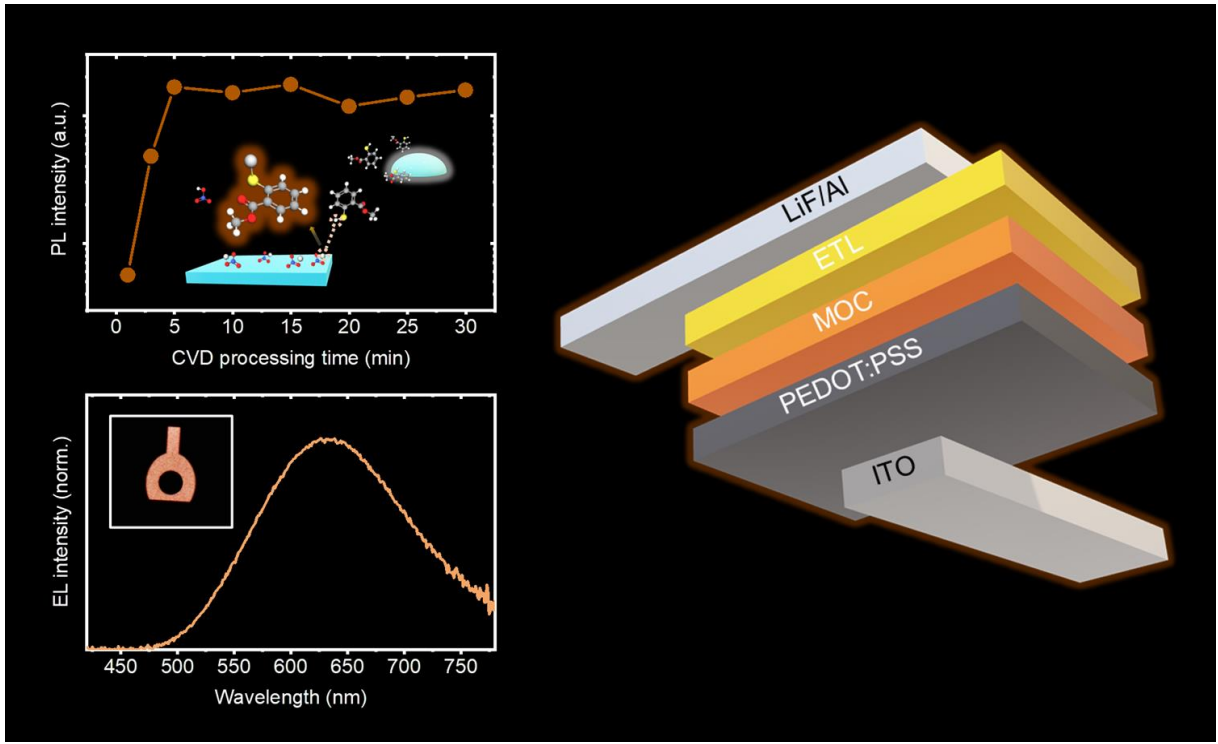

# Rapid Vapor-Assisted Solution Process of Metal-Organic Chalcogenides for High-Performance Light-Emitting Diodes

Sang-Hyun Chin[1,+], Daseul Lee[2,+], Donggyu Lee[1,+], Kwanghyun Chung[1,+], Eunjong Yoo[1], Tong-Il Kim[1], Su Hwan Lee[2], Sang Woo Bae[2], Young-Hoon Kim[2,*], and Yeonjin Yi[1,*]

1. Department of Physics, Yonsei University, 50 Yonsei-ro, Seodaemun-gu, Seoul 03722, Republic of Korea.

2. Department of Energy Engineering, Hanyang University, 222 Wangsimni-ro, Seongdong-gu, Seoul 04763, Republic of Korea.

**\*Corresponding authors**.

E-mail address: younghoonkim@hanyang.ac.kr, (Y.K.), yeonjin@yonsei.ac.kr (Y.Y.)

[+] These authors contributed equally to this work.

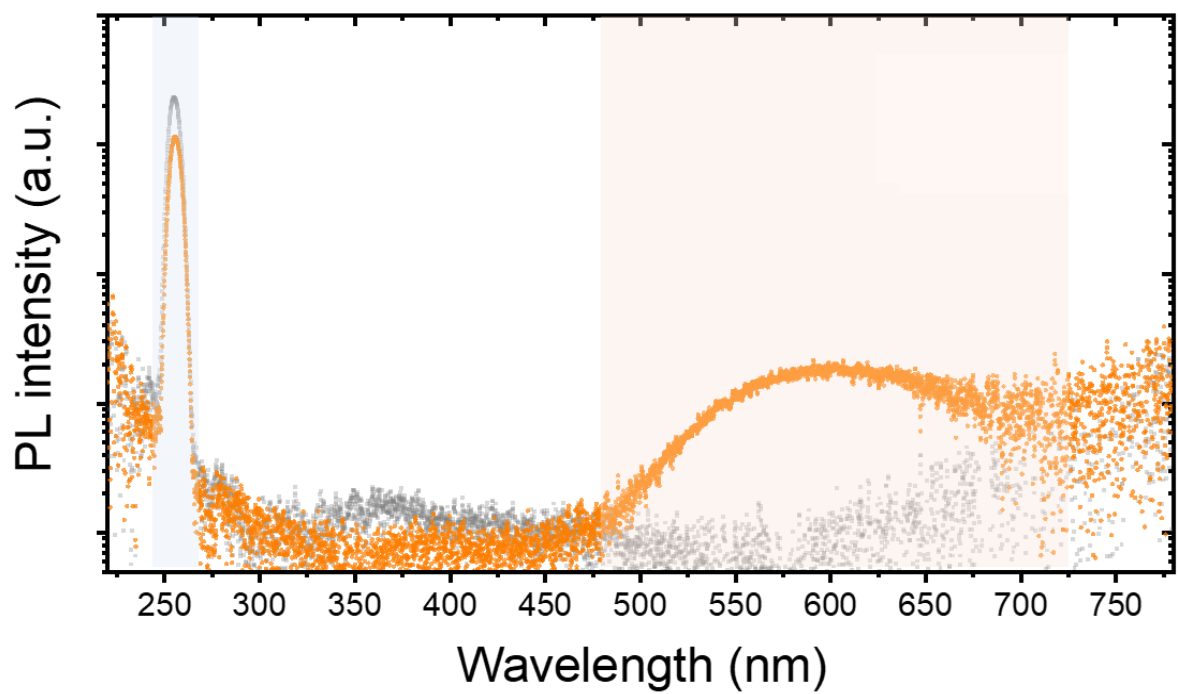

**Figure S1**. Photoluminescence (PL) quantum yield measurement of metal-organic chalcogenide films under 256 nm excitation.

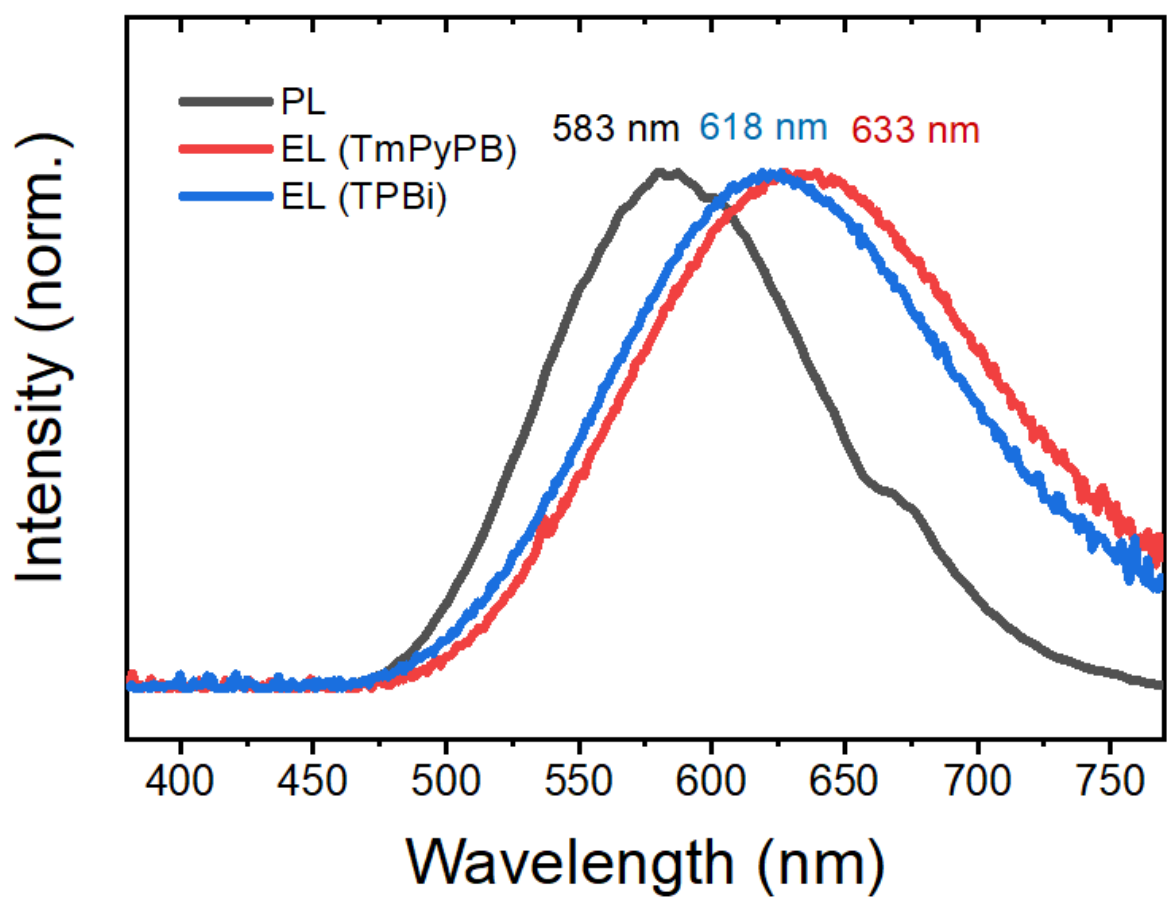

**Figure S2**. Electroluminescent spectra of MOCLEDs.

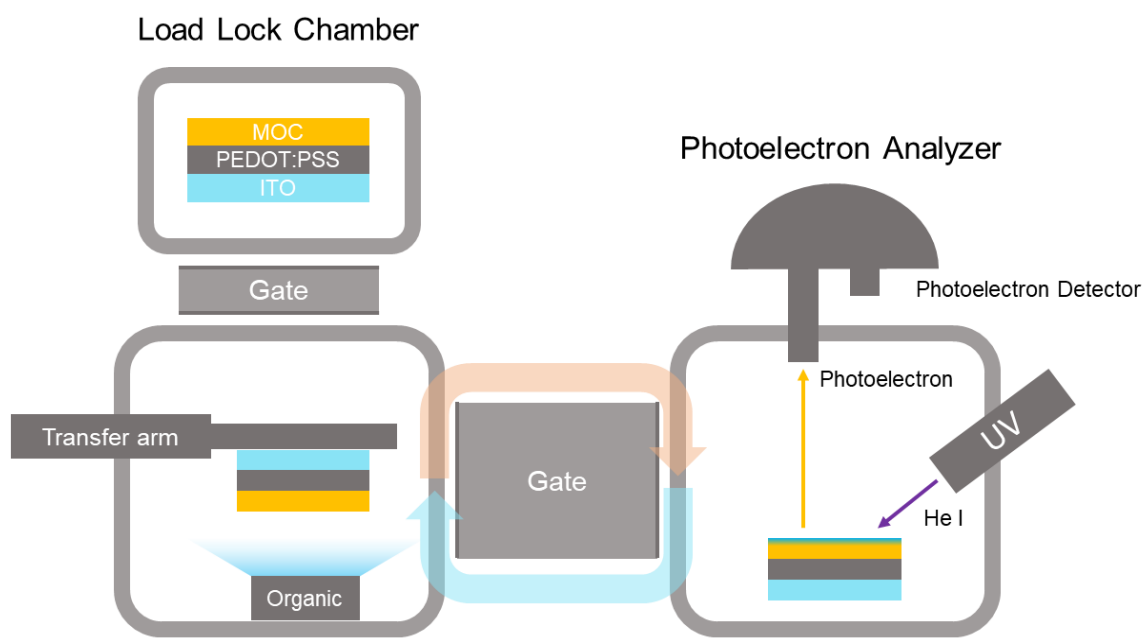

**Figure S3**. Schematic illustration of *in situ* photoelectron spectroscopy.

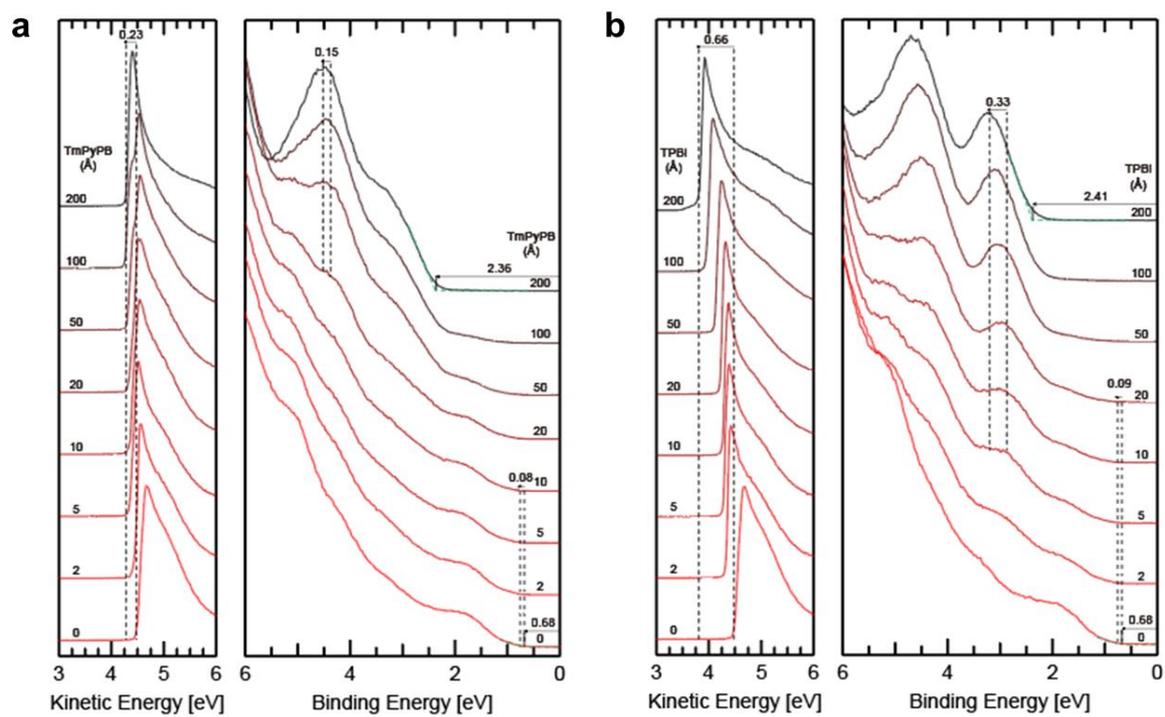

**Figure S4**. Ultraviolet photoelectron spectroscopy (UPS) results of (**a**) MOC-TmPyPB interface and (**b**) MOC-TPBi, presenting secondary cut-off (SECO), valence band (VB) region during the *in-vacuo* study.

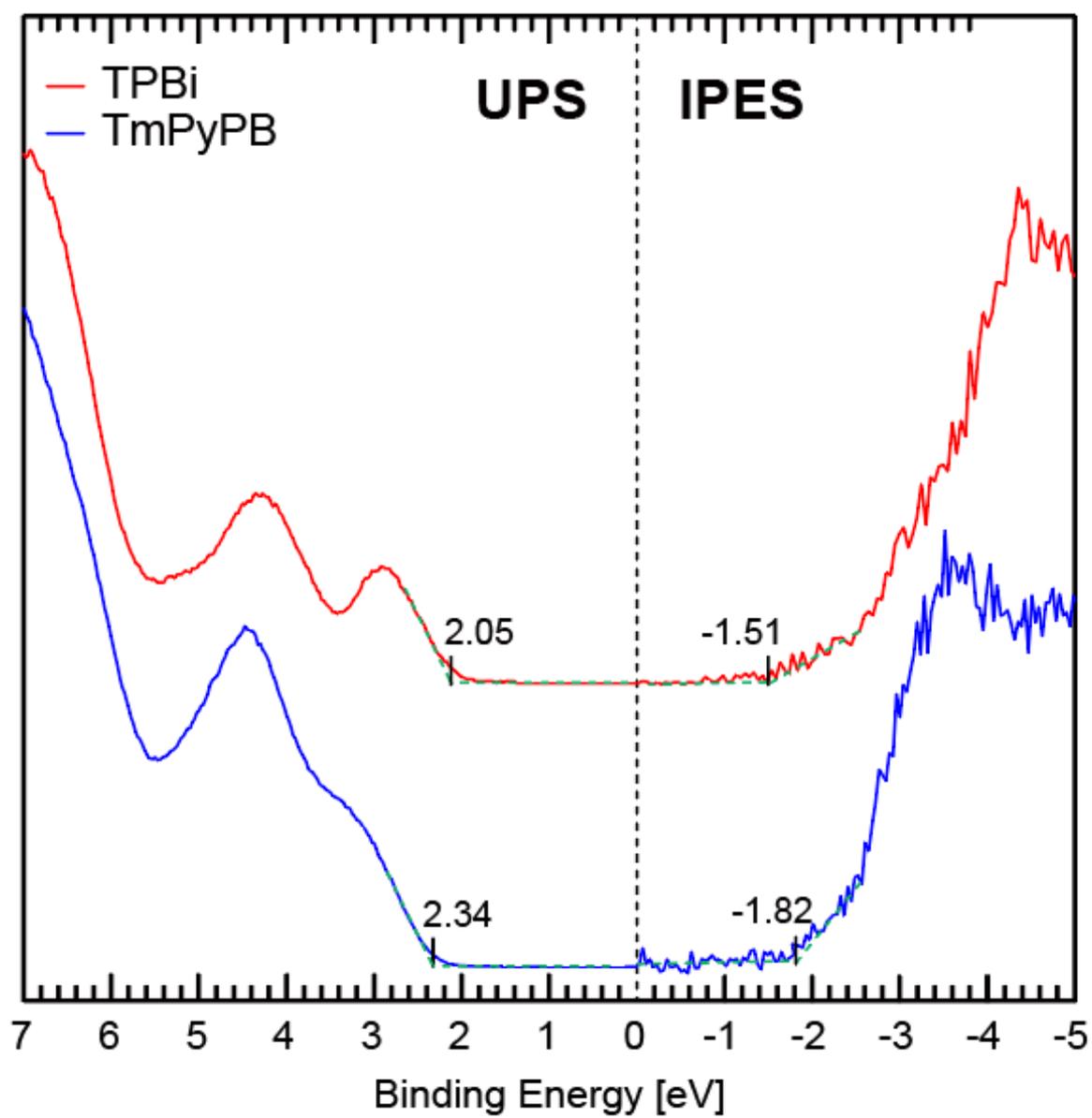

**Figure S5**. UPS-IPES spectra of TmPyPB and TPBi.